\documentclass[runningheads,a4paper]{llncs}

\usepackage{amssymb}
\usepackage{graphicx}
\usepackage{verbatim}

\usepackage{url}
\urldef{\mailsa}\path|{garima.ahuja@students.iiit.ac.in, kamal@iiit.ac.in}|
\newcommand{\keywords}[1]{\par\addvspace\baselineskip
\noindent\keywordname\enspace\ignorespaces#1}

\usepackage{algorithm}
\usepackage{algorithmic}

\begin{document}

\mainmatter  

\title{Crowd Congestion and Stampede Management through Multi Robotic Agents}

%
%
\author{Garima Ahuja \and Kamalakar Karlapalem}
%

\institute{Center for Data Engineering,\\
International Institute of Information Technology Hyderabad, India\\
\mailsa\\}

%
%

\maketitle

\begin{abstract}
Crowd management is a complex, challenging and crucial task. Lack of appropriate management of crowd has, in past, led to many unfortunate stampedes with significant loss of life. To increase the crowd management efficiency, we deploy automated real time detection of stampede prone areas. Then, we use robotic agents to aid the crowd management police in controlling the crowd in these stampede prone areas. While doing so, we aim for minimum interference by robotic agents in our environment. Thereby not disturbing the ambience and aesthetics of the place. We evaluate the effectiveness of our model in dealing with difficult scenarios like emergency evacuation and presence of localized congestion. Lastly, we simulate a multi agent system based on our model and use it to illustrate the utility of robotic agents for detecting and reducing congestion.
\keywords{Crowd Management, Stampede, Congestion, Localized Congestion, Emergency Evacuation}
\end{abstract}

\section{Introduction}

Managing large crowds is a difficult task. Ineffective execution of this task can potentially lead to stampedes. History suggests that crowd management police appointed to prevent such incidents have not been very successful. During a coronation festival in Russia, a police force of 1,800 men failed to control the crowd, leading to loss of 1,389 lives by trampling \cite{khodynka}. Crowd management is a team task, it requires strategic communication to figure out where to lead the crowd. And due to perennial availability of agents, communication between agents is more reliable than communication between humans. Moreover, these incidents are not entirely unexpected, they repeatedly occur at similar occasions like religious gatherings \cite{kumbh,hajj}, music concerts, sports tournaments, etc \cite{hillsborough,who}. Wherever large crowds gather in a relatively small area, risk of stampede is high. And that is why police officers in large numbers are assigned at these places to ensure safe management of crowd. To help the police officers in crowd management, we propose a multi agent based solution. More specifically, in this paper, we deal with the problem of congestion detection and congestion control in large crowds. We assume a two dimensional space on ground where humans can move in any arbitrary direction. Given such a scenario, we propose automated strategies to detect congested areas of the field and an automated strategy for congestion reduction. As demonstrated in our simulations, the agents we propose can successfully detect and control congestion for obedient crowds. The agents can also successfully detect presence of in-obedient crowds and request police intervention in such scenarios.

The organization of the paper is as follows. In section 2, we describe our crowd management model. Following which, we present our congestion detection strategies and address congestion control in sections 3 and 4 respectively. Then, we illustrate the results in section 5. In section 6, we discuss the existing work in crowd dynamics modeling and crowd management. Finally we present conclusion in section 7.

\section{Crowd Management Agents}
We have two kinds of agents for crowd management: Congestion Detecting Agents (CDAs) and Congestion Controlling Agents (CCAs). Congestion detecting agents inspect parts of the field and report presence or absence of congestion in them. Congestion controlling agents take pro-active measures to reduce congestion in a congested area.

Congestion Detecting Agents (CDAs) are computing machines present off-site. They receive location coordinates of humans present inside the field. (We do not delve into the details of how we receive the location coordinates, possible ways to obtain positional information are GPS or RFID). CDAs then use this information to check for congestion.

Congestion Controlling Agents (CCAs) on the other hand are flying robots (like droids). They have a spotlight and a speaker attached to them to allow them to illuminate the areas below them and to announce instructions. CCAs receive instructions from CDAs and execute them to achieve congestion reduction.

Our CDAs never enter the site. This helps in accomplishing minimum interference by robotic agents in our human environment. We do not send any robotic agent inside the field unless there is a stampede risk; which is when, it becomes absolutely necessary to do so.
In the following sections, we present how we achieve congestion detection and congestion control using the agents described.

\section{Congestion Detection}
For computational ease and efficiency, we divide our field into smaller rectangular areas (henceforth called grids). The division allows multiple CDAs to parallely inspect grids. 

We fork as many CDAs as there are grids to parallely inspect the grids and check for presence of congestion in them. For the grids for which presence of congestion is reported, we find the extent and boundaries of the congested area using the breadth first search algorithm Algorithm 1. 

\begin{algorithm}
    \caption{Given a grid id as input, return grid ids of the neighbouring congested grids.}
    \begin{algorithmic}[1]
        \STATE{$Given: gridId$}
        \STATE{list $congestedGridIds \leftarrow emptyList$}    
        \STATE{list $toBeCheckedGridIds \leftarrow emptyList$}
        \STATE{$toBeCheckedGridIds.add(gridId$)}
        \WHILE{$toBeCheckedGridIds \neq empty$}
        \STATE{$currentGridId \leftarrow toBeCheckedGridIds.getFirstElement()$}
        \STATE{/* }
        \STATE{Strategies for checkForCongestion() function are described in subsections of section 3.}
        \STATE{*/}
        \IF{$checkForCongestion(currentGridId) == true$}
        \STATE{//congestion present}
        \STATE{$congestedGridIds.add(currentGridId$)}
        \STATE{list $neighbouringGridIds \leftarrow getFourOverlappingGridIds(currentGridId)$}
        \FOR{each $neighbouringGridId$ in $neighbouringGridIds$}
        \STATE{/*}
        \STATE{ One of the overlapping grids of a child grid would be the parent congested grid itself, gridId of which would be present in congestedGridIds. Since we have already checked the parent congested grid, we do not add it to the toBeCheckedGridIds list.}
        \STATE{*/}
        \IF{$neighbouringGridId$ not present in $congestedGridIds$}
        \STATE{$toBeCheckedGridIds.add(neighbouringGridId$)}
        \ENDIF
        \ENDFOR
        \ENDIF
        \STATE{/*}
        \STATE{The first element of the toBeCheckedGridIds list is currentGridId. We finished inspecting the grid corresponding to the currentGridId, so we remove it from the list.} 
        \STATE{*/}
        \STATE{$toBeCheckedGridIds.removeFirstElement()$}   
        \ENDWHILE
        \RETURN $congestedGrids$
    \end{algorithmic}
\end{algorithm}

The CCAs need to know the boundaries of the congested areas to resolve congestions. We discuss the details of congestion control in section 4.

In the following subsections, we describe the different strategies that could be used for congestion detection in a grid.

\subsection{Naive Strategy}
We can estimate the capacity of a grid based on its area. A CDA would then count the number of humans present inside the grid and report presence of congestion if this number is more than the capacity of the grid. Otherwise, the CDA would report absence of congestion.

\subsection{Free Flow Strategy}
CDA counts the number of humans who entered or left the grid in one unit time. If the number of humans who left the grid is more than the number of humans who entered, we say that there is free flow in the grid and report absence of congestion; otherwise we report presence of congestion.

\subsection{Trapped Humans Strategy}
We report presence of congestion in the grid when there are one or more trapped humans present. A human is trapped when he has more than minN neighbours in his neighbourhood where the neighbourhood is defined as a circle of radius r around a human. Figure 1 shows one such scenario. This strategy is not computationally efficient as it involves inspecting neighbourhood of every human present in the grid. However, search space can be decreased using Locality Sensitive Hashing (LSH), thereby improving the performance.

\begin{figure}
\centering
\includegraphics[height=3cm]{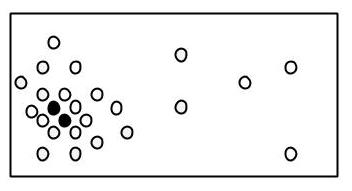}
\caption{Trapped humans are represented by filled circles.}
\end{figure}

\subsection{A Macro Micro Strategy}
The strategies suggested so far use only the location coordinates of humans present inside the field for congestion detection. We can however use a human's previous and current positions to determine the direction he/she is moving in. We can also record the time stamps as and when we receive the location coordinates and use it to determine his/her speed.

Another important piece of information we have not considered in the other strategies is the human behaviour. Humans tend to visit places in groups of families or friends \cite{kota}. In a particular group, all the humans tend to be close to each other and move in the same direction. Their direction of movement also depends on their inherent tendency to avoid collisions.

We now propose a macro micro strategy for congestion detection which uses all the information we have not been utilizing till now. We divide the humans into groups where a group is defined as a connected group of humans moving in the same direction. The congestion detection then proceeds as follows:

(i) At the macro or inter-group level, we count the number of groups that are moving towards a fixed point. We report presence of congestion if this number is greater than or equal to three. (Our simulations show that when the number of conflicting directions is two, the human tendency to avoid collisions is ensuring hassle free movement.) 

(ii) At the micro or intra-group level, we further divide the groups into connected sub-groups of humans moving with similar speeds. We report presence of congestion if speed of a sub-group is more than the speed of the sub-group moving in front of it (front is along the direction of movement of the group).

(iii) If both the conditions stated above do not hold, we report absence of congestion.

\section{Congestion Control}
Our simulation results (presented in the next section) show that the macro micro strategy performs better than the other strategies for congestion detection. We therefore use it and, based on it, devise a strategy for congestion control.

Recall that the CDA not just reports presence or absence of congestion in a grid but also finds out the boundaries of the congested area. Thus, the CDA knows where to lead the humans present in the congested area, i.e. out of the congestion boundary. The CDA calls a CCA and then they work together to achieve congestion control. One of the following cases will arise:

(i) There are no conflicting directions (intersecting directions) but the speed of a sub-group is greater than the speed of the sub-group moving in front of it (front is along the direction of movement of the group). Recall that the CCA has a spotlight attached to it to illuminate the area below. The CCA starts moving above the sub-group while shining its spotlight on the sub-group and playing a pre-recorded message requesting the humans to slow down.

(ii) There are conflicting directions but at least one of the group has space in front to proceed (see Figure 2). The CCA then positions itself above that group and leads the group in its target direction while requesting the other groups to wait for it to return. The group being led is requested to follow the spotlight and maintain the speed of the spotlight. After the group is led out of the congested area, the new scenario could conform to case (ii) or case (iii) and is handled accordingly recursively till congestion is resolved.

\begin{figure}
\centering
\includegraphics[height=2.5cm]{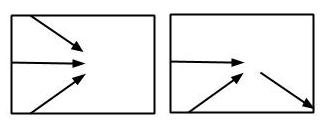}
\caption{When there is space in front of at least one of the groups, that group is led while the others wait.}
\end{figure}

(iii) There are conflicting directions and none of the groups have space in front to proceed (see Figure 3). Using the pre-recorded messages and the spotlight, the CCA leads one of the group through a semi circular path around the congested area while requesting the other groups to wait for it to return. The center of the semi circular path is given by (1) and (2) and the radius is given by (3).

\begin{equation}
  center_{x}= \{(a_{x}+b_{x})/2|dist(a,b) \geq dist(i,j)\forall i,j \in humansInCongestedArea \} \\
\end{equation}

\begin{equation}
  center_{y}= \{(a_{y}+b_{y})/2|dist(a,b) \geq dist(i,j)\forall i,j \in humansInCongestedArea \} 
\end{equation}

\begin{equation}
  radius = \{dist(a,b)/2|dist(a,b) \geq dist(i,j)\forall i,j \in humansInCongestedArea \} 
\end{equation}

\begin{figure}
\centering
\includegraphics[height=2.5cm]{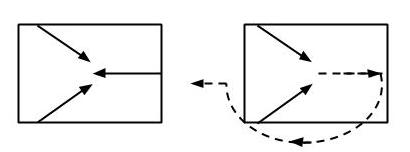}
\caption{When there is no space in front of any of the groups, one of the group is led through a semi circular path around the congested area while the other groups wait.}
\end{figure}

After the group is led out of the congested area, the new scenario could conform to case (ii) or case (iii) and is handled accordingly recursively till congestion is resolved.

Notice in Figure 3 that the groups finally end up moving in the direction they originally intended to. Only difference being that the CCAs made them move around the congested area while they were trying to move through the congested area.

Also, note that the isVacant() function used in the congestion boundary detection algorithm (presented in section 3) would use results from naive strategy, not macro micro strategy. The reason being that the CCAs need to know where there is empty space so that they can lead the conflicting groups accordingly. Macro Micro strategy would report absence of congestion for densely populated grids too as long as there are no conflicting directions present and therefore it cannot be used for the isVacant() function. 

The environment we are dealing with is non-deterministic. The outcome of the congestion control strategy would depend on the obedience of the crowd. Therefore, it is crucial to take feedback from the environment and act accordingly. To that end, the CDA keeps recalculating humans' speeds and directions to calculate the percentage of disobedience. A human is disobedient if he/she does not have the speed or direction he/she is being instructed to have. If the percentage of disobedience is high, the CDA requests the police in charge to take over and manage the crowd.

\section{Results}
We use Helbing and Molnar's social force model \cite{helbing} for crowd simulation. The motion of each human is governed by the summation of all the forces exerted on and by the human. The following forces help us in modeling group behavior:

(i) \textbf{Intent}: A human exerts force in the direction he wants to move in.

(ii) \textbf{Cohesion}: A human exerts force so as to remain close to his group.

(iii) \textbf{Coherency}: A human exerts force so as to walk in the same direction as his group.

(iv) \textbf{Momentum}: Inertial force.

(v) \textbf{Avoidance}: A human exerts force to avoid colliding with other humans and obstacles.

The first scenario we consider is an emergency evacuation scenario. To simulate that, we add a force towards the exit door to the above forces. In an evacuation scenario, we do not want our congestion controlling agents to interrupt the evacuation if the humans are moving towards the exit door in a calm manner. So we want absence of congestion to be reported if the movement of the crowd is not unruly. Figure 4 shows such a scenario where the humans are moving in an orderly manner towards the exit gate while avoiding obstacles. Naive strategy, free flow strategy and trapped humans strategy all fail in the evacuation scenario because they cannot distinguish between a calm crowd and an unruly one. However, macro micro strategy recognizes the gradually changing direction as one direction and reports absence of congestion if speed conflicts are absent. Therefore, we will not unnecessarily interrupt evacuation if the crowd is calm (speed conflicts absent). However as soon as the crowd becomes unruly, speed conflicts will be detected and remedial measures will be taken. Hence, our crowd managing model is able to successfully handle the emergency evacuation scenario without any unnecessary interruptions.

\begin{figure}
\centering
\includegraphics[height=4cm]{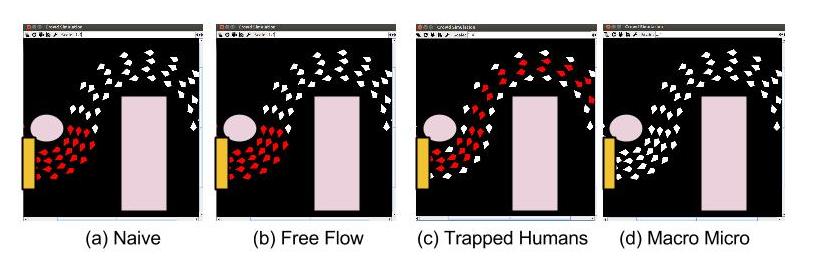}
\caption{Outputs of different congestion detection strategies when humans are moving calmly towards the golden exit gate while avoiding obstacles.  Direction of the compass represents direction of the human. Red compasses represent congestion. }
\end{figure}

We now consider a localized congestion scenario i.e. a scenario where congestion is localized to a relatively small area (in this case a grid). To simulate this scenario, to the above listed forces we add a force directed towards a fixed point. Results of various congestion detection strategies for localized congestion scenario is shown in Figure 5. Naive strategy, free flow strategy and trapped humans strategy again fail in this scenario because the grid is sparsely populated and number of humans entering the grid is zero. However, macro micro strategy is able to identify the root cause behind congestion which is the conflicting directions and reports congestion successfully. 

\begin{figure}
\centering
\includegraphics[height=4cm, width=12cm]{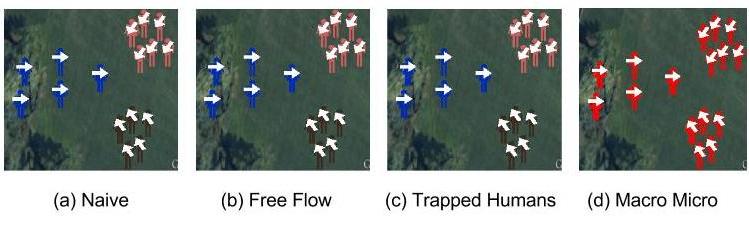}
\caption{Outputs of different congestion detection strategies for the localized congestion scenario.  Direction of the arrow represents direction of the human. Red humans represent congestion.}
\end{figure}

Now that we have shown that the macro micro strategy works better than other strategies for congestion detection, we test our congestion control strategy for the above stated localized congestion scenario. The results are presented in Figure 6. Figure 6(a) shows the present scenario. Figure 6(b) shows the blue group being led on a semi circular path. Figure 6(c) shows the brown group being led on a straight path. Figure 6(d) shows the scenario after congestion is successfully resolved.

\begin{figure}
\centering
\includegraphics[height=4cm, width=12cm]{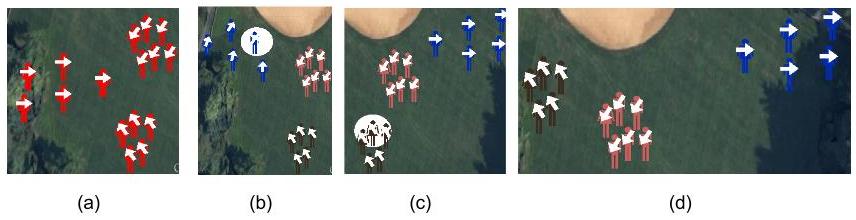}
\caption{Output of macro micro congestion control for the localized congestion scenario. The present scenario is shown in (a), (b) shows the blue group being led on a semi circular path, (c) shows the brown group being led on a straight path and (d) shows the scenario after congestion is successfully resolved.}
\end{figure}

Finally, we simulate humans on a central park background and test our congestion detection and congestion control strategies. The results are presented in Figure 7. 

\begin{figure}
\centering
\includegraphics[height=4cm, width=12cm]{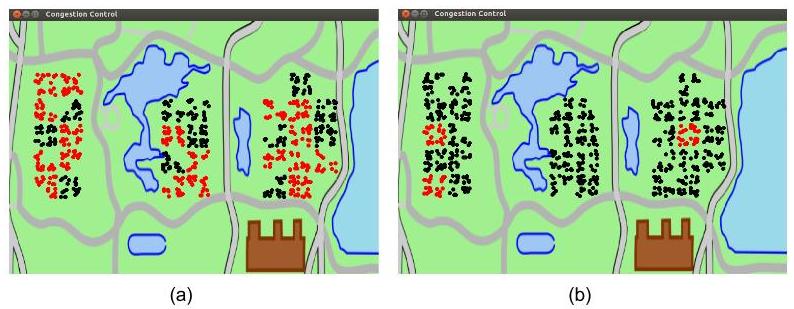}
\caption{Situation before(a) and after(b) congestion control. Red dots represent humans experiencing congestion.}
\end{figure}

The time taken to resolve congestion for different magnitudes of congestion is tabulated in Table 1.

\begin{table*}
\centering
\caption{Time taken to resolve congestion}
\begin{tabular}{|p{1cm}|p{2.5cm}|p{2.5cm}|p{2.5cm}|p{2.5cm}|p{2.5cm}|} \hline
No.&Number of humans in the congested area&Number of conflicting groups in the congested area&Maximum walking speed of a human in metres per second&Time taken to resolve congestion in seconds
\\ \hline
1&6&3&1.4&2.2\\ \hline
2&12&3&1.2&10.9\\ \hline
3&19&3&1.4&24.3\\ \hline
4&28&5&1.3&97.6\\ \hline
5&34&6&1.3&134.0\\ \hline
6&41&8&1.2&247.8\\ \hline
\end{tabular}
\end{table*}

Remember that we pointed out that the groups finally end up moving in the direction they originally intended to. Only difference being that the CCAs made them move around the congested area while they were trying to move through the congested area. Since we are not making the humans move in a direction they did not intend to, we cannot assure complete absence of congestion. Direction conflicts will inevitably arise unless we force the humans to move in a direction they do not want to move in (see Figure 7(b)). Nevertheless, our congestion control strategy resolves congestion as soon as it is detected and keeps on doing so to ensure hassle free movement of the crowd.

\section{Related Work}
There has been a lot of research in trying to model crowd dynamics and behaviour. There are cellular automata based models \cite{dijkstra,hamagami,kirchner} and agent-based models \cite{murakami,osaragi,toyama}. Helbing and Molnar \cite{helbing} model is a social force model that simulates motion of each individual in the crowd under the
influence of the forces exerted on the individual by other individuals and inanimate objects. An individual can exert two types of forces, physical and social. The social forces represent the intention of a human to move in a chosen direction and avoid collision. The resultant motion of the human is governed by sum of all forces exerted by the human and on the human. We used this model to simulate human agents in our experiments.

These crowd behaviour models help in increasing our understanding and provide a framework to work towards crowd management models. Helbing, Johansson and Al-Abideen studied the dynamics of crowd disasters \cite{helbing_johansson}. They analyzed the videos of crowd disasters during Hajj
and studied crowd response and behaviour. Helbing and Mukerji analyzed the Love Parade disaster \cite{helbing_mukerji}. Sime \cite{sime} proposed that early warning would help in efficient crowd management during disasters. Kolli and Karlapalem \cite{sindhu} presented crowd management strategies in long queues, a
one-dimensional and unidirectional scenario. However, the problem of crowd management in a two-dimensional scenario using robotic agents proposed in this paper has not been studied.

\section{Conclusion}
In this paper, we presented a multi agent based solution to crowd management. Our results show that the agents are able to detect the presence of congestion as well as take remedial actions even in difficult scenarios like localized congestion and emergency evacuation. Thus the robotic agents help the police force in ensuring a congestion free movement of the crowd. Future work may look into how to handle congestion in a three dimensional scenario where say there are stairways present. Such a scenario is more dangerous as the chances of stumbling and falling increase and therefore it needs to be addressed.

\bibliographystyle{abbrv}
\bibliography{congestion_control}

\begin{thebibliography}{10}

\bibitem{kumbh}
Allahabad stampede kills 36 kumbh mela pilgrims.
\newblock
  http://in.reuters.com/article/2013/02/11/kumbh-mela-stampede-allahabad-update-idINDEE91907I20130211.

\bibitem{hillsborough}
Hillsborough disaster and its aftermath.
\newblock http://www.bbc.com/news/uk-19545126.

\bibitem{hajj}
A history of hajj tragedies.
\newblock http://www.theguardian.com/world/2006/jan/13/saudiarabia.

\bibitem{khodynka}
Khodynka tragedy.
\newblock http://worldhistoryproject.org/1896/5/18/khodynka-tragedy.

\bibitem{who}
The who concert disaster.
\newblock http://en.wikipedia.org/wiki/The\_Who\_concert\_disaster.

\bibitem{dijkstra}
J.~Dijkstra, H.~J. Timmermans, and A.~Jessurun.
\newblock A multi-agent cellular automata system for visualising simulated
  pedestrian activity.
\newblock In {\em Theory and Practical Issues on Cellular Automata}, pages
  29--36. Springer, 2001.

\bibitem{hamagami}
T.~Hamagami and H.~Hirata.
\newblock Method of crowd simulation by using multiagent on cellular automata.
\newblock In {\em Intelligent Agent Technology, 2003. IAT 2003. IEEE/WIC
  International Conference on}, pages 46--52. IEEE, 2003.

\bibitem{helbing_johansson}
D.~Helbing, A.~Johansson, and H.~Z. Al-Abideen.
\newblock Dynamics of crowd disasters: An empirical study.
\newblock {\em Physical review E}, 75(4):046109, 2007.

\bibitem{helbing}
D.~Helbing and P.~Molnar.
\newblock Social force model for pedestrian dynamics.
\newblock {\em Physical review E}, 51(5):4282, 1995.

\bibitem{helbing_mukerji}
D.~Helbing and P.~Mukerji.
\newblock Crowd disasters as systemic failures: analysis of the love parade
  disaster.
\newblock {\em EPJ Data Science}, 1(1):1--40, 2012.

\bibitem{kirchner}
A.~Kirchner and A.~Schadschneider.
\newblock Simulation of evacuation processes using a bionics-inspired cellular
  automaton model for pedestrian dynamics.
\newblock {\em Physica A: Statistical Mechanics and its Applications},
  312(1):260--276, 2002.

\bibitem{sindhu}
S.~Kolli and K.~Karlapalem.
\newblock Mama: multi-agent management of crowds to avoid stampedes in long
  queues.
\newblock In {\em Proceedings of the 2013 international conference on
  Autonomous agents and multi-agent systems}, pages 1203--1204. International
  Foundation for Autonomous Agents and Multiagent Systems, 2013.

\bibitem{kota}
R.~Kota, V.~Bansal, and K.~Karlapalem.
\newblock System issues in crowd simulation using massively multi-agent
  systems.
\newblock In {\em Workshop on Massively Multi Agent Systems}, pages 251--257,
  2006.

\bibitem{murakami}
Y.~Murakami, K.~Minami, T.~Kawasoe, and T.~Ishida.
\newblock Multi-agent simulation for crisis management.
\newblock In {\em Knowledge Media Networking, 2002. Proceedings. IEEE Workshop
  on}, pages 135--139. IEEE, 2002.

\bibitem{osaragi}
T.~Osaragi.
\newblock Modeling of pedestrian behavior and its applications to spatial
  evaluation.
\newblock In {\em Proceedings of the Third International Joint Conference on
  Autonomous Agents and Multiagent Systems-Volume 2}, pages 836--843. IEEE
  Computer Society, 2004.

\bibitem{sime}
J.~D. Sime.
\newblock Crowd facilities, management and communications in disasters.
\newblock {\em Facilities}, 17(9/10):313--324, 1999.

\bibitem{toyama}
M.~C. Toyama, A.~L. Bazzan, and R.~Da~Silva.
\newblock An agent-based simulation of pedestrian dynamics: from lane formation
  to auditorium evacuation.
\newblock In {\em Proceedings of the fifth international joint conference on
  Autonomous agents and multiagent systems}, pages 108--110. ACM, 2006.

\end{thebibliography}

\end{document}